\documentclass{aa}
\usepackage[english]{babel}
\usepackage[varg]{txfonts}
\usepackage[squaren,Gray]{SIunits}

\usepackage[usenames,dvipsnames]{xcolor}

\graphicspath{{./}{figures/}}

\usepackage{epsfig}
\usepackage{graphicx,natbib}
\usepackage{amsmath,amssymb}
\usepackage{hyperref}
\usepackage{wasysym}
\usepackage{natbib}
\usepackage{ulem}
\usepackage{cancel}
\usepackage{subfigure}

\hypersetup{
    colorlinks=true,
    citecolor=blue,
    linkcolor=red,
    urlcolor=black
    }

\begin{document}

\def\nat{Nature }
\def\apj{ApJ}
\def\apjs{ApJS} 
\def\apjl{ApJ} 
\def\apss{Ap. Sp. Sci.}
\def\aap{A\&A}
\def\mnras{MNRAS}
\def\jgr{JGR}

\def\nh{n_\mathrm{H}}
\def\nho{n_{\mathrm{H}0}}
\def\cc{\mathrm{cm}^{-3}}
\def\ni{n_\mathrm{i}}
\def\ne{n_\mathrm{e}}
\def\svie{\langle \sigma v \rangle_\mathrm{ie}}
\def\depsdpsi{\frac{\mathrm{d}\epsilon}{\mathrm{d}\psi}}
\def\dnidpsi{\frac{\mathrm{d}\ni}{\mathrm{d}\psi}}

\title{Fast methods for tracking grain coagulation and ionization. I. Analytic derivation}
\titlerunning{}

\author{P. Marchand \inst{1}, V. Guillet \inst{2,3}, U. Lebreuilly \inst{4,5}, M.-M. Mac Low \inst{1,6}}

\institute{Department of Astrophysics, American Museum of Natural History, Central Park West at 79th Street, New York, NY 10024, USA
\and Universit\'e Paris-Saclay, CNRS, Institut d’astrophysique spatiale, 91405, Orsay, France
\and Laboratoire Univers et Particules de Montpellier, Universit\'e de Montpellier, CNRS/IN2P3, CC 72, Place Eug\`ene Bataillon, 34095 Montpellier Cedex 5, France
\and Centre de Recherche Astrophysique de Lyon, ENS Lyon, 46 All\'ee d'Italie, 69007 Lyon, France
\and AIM, CEA, CNRS, Université Paris-Saclay, Université Paris Diderot, Sorbonne Paris Cité, F-91191 Gif-sur-Yvette, France
\and Center for Computational Astrophysics, Flatiron Institute, 162 Fifth Avenue, New York, NY 10010, USA}

\authorrunning{P. Marchand et~al.}

\date{}

\abstract{Dust grains play a major role in many astrophysical contexts. They affect the chemical, magnetic, dynamical, and optical properties of their environment, from galaxies down to the interstellar medium, star-forming regions, and protoplanetary disks. Their coagulation leads to
  shifts in their size distribution and ultimately
  to the formation of planets. However, although the coagulation process is reasonably uncomplicated to numerically implement by itself, it is difficult to couple it with multidimensional hydrodynamics numerical simulations because of its high computational cost. We propose here a simple method for tracking the coagulation of grains at far lower cost.
   Given an initial grain size distribution, the state of the distribution at time t is solely determined by the value of a single variable integrated along the trajectory, independently of the specific path taken by the grains. Although this method cannot account for processes other than coagulation, it is mathematically exact, fast, inexpensive, and can be used to evaluate the effect of grain coagulation in most astrophysical contexts. It is applicable to all coagulation kernels in which local physical conditions and grain properties can be separated. We also describe another method for calculating the average electric charge of grains and the density of ions and electrons in environments that are shielded from radiation fields, given the density and temperature of the gas, the cosmic-ray ionization rate, and the average mass of the ions. The equations we provide are fast to integrate numerically and can be used in multidimensional numerical simulations to self-consistently calculate on the fly the local resistivities that are required to model nonideal magnetohydrodynamics.}


\maketitle

\section{Introduction}

Grains are a fundamental component of the universe. These molecular aggregates represent 1\% of the mass of the interstellar medium \citep[ISM;][]{mathis,2001ApJ...548..296W} and play a major role in the chemistry, dynamics, and thermodynamics of their environment in the ISM and during star and planet formation.
Grain surfaces catalyze chemical reactions of gaseous species \citep{2015A&A...576A..49H}, and they can hold several electric charges \citep{DraineSutin}. They therefore affect the chemical and ionization equilibrium of the gas and the nature of the ionized species, which control the resisitivities producing nonideal magnetohydrodynamics (MHD) effects \citep{2016A&A...592A..18M} and the regulation of angular momentum during star formation \citep{2016MNRAS.460.2050Z,2018MNRAS.473.4868Z,Marchand2020}. In protostellar environments, grains are the main source of opacity and significantly affect the thermal properties of protoplanetary disks \citep{1997A&A...325..569S} that lead to the formation of the first hydrostatic core \citep{Larson1969}. Additionally, their optical properties, including their absorption of infrared radiation, is a major factor to account for in observations \citep{Semenov2003}.

In the diffuse ISM, the dust grain population is well described by the Mathis-Rumpl-Nordsieck (MRN) size distribution \citep{mathis}. This distribution is often assumed to be preserved in the denser parts of the ISM, although it can in principle evolve with time as grains grow by accretion or by sticking together through collisions: coagulation \citep{1982A&A...114..245T,1991A&A...251..587R,1993ApJ...407..806C,2009A&A...502..845O,2017A&A...603A.105D}.  These effects can reduce or remove smaller size grains in the densest regions of the ISM, which is consistent with observational evidence \citep{1989ApJ...345..245C,1993AJ....105.1010V}. In star-forming contexts, grains lying in protoplanetary disks coagulate at a faster rate than in the ISM and grow from submicron sizes to micrometer and millimeter sizes. Eventually, these grains become the seeds of planet formation. 

Modeling and observing accurate dust size distributions, however, is a difficult challenge. Several studies have accounted for the coagulation of grains and its astrophysical implications \citep{2009A&A...502..845O,2017MNRAS.467..699H,2020A&A...643A..17G}, often using complex chemical or microphysics codes to compute accurate results \citep{1993ApJ...407..806C,1997AdSpR..20.1595P,1997ApJ...480..647D}, or assuming a size distribution that emulates the coagulation \citep{2016MNRAS.460.2050Z,2020ApJ...896..158T,Marchand2020}. Here we propose a new method for simulating the coagulation of grains when the grain velocity only depends on the local physical conditions (density, temperature, magnetic field, etc.). We apply this to the commonly used collision kernel derived by \citet{2007A&A...466..413O}, and find a new expression of the well-known Smoluchowski coagulation equation \citep{1916ZPhy...17..557S}. The coagulated distribution is solely determined by an initial size distribution and the value of a quantity $\chi$, which is an integral of physical conditions along the path of the grains. The coagulation process is therefore reduced to a 1D problem because the evolution of the size distribution only depends on $\chi$ and not on the specific path taken.
Promising methods have been developed to accelerate the solving of the Smoluchowski equation, such as implicit and semi-implicit integration schemes \citep{2008ApJ...682..515E} or Garlekin schemes \citep{2021MNRAS.501.4298L}. However, the main advantage of our method is that the Smoluchowski equation does not have to be solved during hydrodynamical simulations.

Nonideal MHD effects play a major role in the regulation of the
angular momentum and the magnetic flux during star formation
\citep{Machida_etal06,MellonLi2009,2015ApJ...801..117T,2016MNRAS.457.1037W,Marchand2020},
but computing the resistivities usually requires intensive computations by specialized chemistry codes \citep{KunzMouschovias2009,2016A&A...592A..18M,2016PASA...33...41W,2019MNRAS.484.2119K}. We propose here a fast and mathematically accurate method based on the model of \citet{DraineSutin} to compute the density of ions, electrons, and the average electric charge of grains, knowing the average ion mass, the density, the temperature, the ionization rate from cosmic rays (CR) for an arbitrary grain size distribution. 

Both methods can be used in conjunction to simply and rapidly calculate the dust size distribution and magnetic resistivities in a self-consistent way. They can be implemented in nonideal magnetohydrodynamical simulations from the scales of molecular clouds down to collapsing protostellar cores.

Our next papers will detail some applications of these methods, while this paper focuses on their derivation. In Sect.~\ref{SecDeriv} we derive the new form of the Smoluchowski equation, and we detail the calculation of the ionization in Sect.~\ref{SecIonization}. Caveats are discussed in Sect.~\ref{SecCaveats}, and conclusions are given in Sect.~\ref{SecConclusion}.

\section{Grain coagulation} \label{SecDeriv}   

The time evolution of the distribution of coagulating grains is classically described by the \citet{1916ZPhy...17..557S} equation \citep[also see][]{1988A&A...195..183M}. In an environment with varying density, the change in time $t$ of the mass density $\rho(m,t)$ of grains with mass $m$ is

\begin{equation}
\begin{aligned}\label{EqSmolu}
  \frac{\mathrm{d}\rho(m,t)}{\mathrm{d}t} = &- \int_0^\infty m K(m,m') n(m,t)n(m',t)\mathrm{d}m'\\
                          &+ \frac{1}{2} \int_0^m m K(m-m',m') n(m-m',t)n(m',t)\mathrm{d}m'\\
                          &+ \frac{\rho(m,t)}{n_\mathrm{H}}\frac{\mathrm{d}n_\mathrm{H}}{\mathrm{d}t},
\end{aligned}
\end{equation}
with $K(m,m')$ the coagulation kernel between grains of mass $m$ and $m'$, $n(m,t)$ the number density of grains of mass $m$ at time $t,$ and $n_\mathrm{H}$ the number density of gas.

\subsection{General form}\label{SecGeneralform}

We assume a kernel in the form

\begin{equation}
  K(m,m') = C g_\mathrm{local} h(m,m'),
\end{equation}
where $C$ is a constant, $g_\mathrm{local}$ is a function of local physical conditions ($\nh$, $T$, $B$, etc.), and $h(m,m')$\footnote{Since $m$ corresponds to a unique $a$ and vice versa, we do not distinguish between them in function arguments.} a function depending only on the grains properties.
The Smoluchowski equation can be rewritten using the functions just defined as

\begin{equation}
\begin{aligned}\label{Smolu2}
  \frac{\mathrm{d}\rho(m,t)}{\mathrm{d}t} = &mC g_\mathrm{local}\left[ - \int_0^\infty h(m,m') n(m,t)n(m',t)\mathrm{d}m' \right.\\
                                            &\left. + \frac{1}{2} \int_0^m h(m-m',m') n(m-m',t)n(m',t)\mathrm{d}m' \right]\\
                                            &+\frac{\rho(m,t)}{n_\mathrm{H}}\frac{\mathrm{d}n_\mathrm{H}}{\mathrm{d}t}.
\end{aligned}
\end{equation}
If $X(a,t)$ is the fraction of grains\footnote{or equivalently, the   dust-to-gas ratio as a function of size} of size $a$ in the gas at time $t$, we have $n(a)=n_\mathrm{H}X(a,t)$, and $\rho(m,t)=mn_\mathrm{H}X(a,t)$. Hence 

\begin{equation}
  \frac{\mathrm{d}\rho(m,t)}{\mathrm{d}t} = m\left[ n_\mathrm{H}\frac{\mathrm{d}X(a,t)}{\mathrm{d}t}+ X(a,t)\frac{\mathrm{d}n_\mathrm{H}}{\mathrm{d}t} \right].
\end{equation}
The second term is equivalent to the last term of the Smoluchowski equation $ (\rho(m,t)/n_\mathrm{H})(\mathrm{d}n_\mathrm{H}/\mathrm{d}t)$, so they cancel out. Equation~(\ref{Smolu2}) thus becomes

\begin{equation}
  \frac{\mathrm{d}X(a,t)}{\mathrm{d}t} = C g_\mathrm{local} n_\mathrm{H} I(a,X,t),
\end{equation}
where

\begin{equation}
\begin{aligned}
  I(a,X,t)=& - \int_0^\infty h(m,m') X(m,t)X(m',t)\mathrm{d}m'\\
           & + \frac{1}{2} \int_0^m h(m-m',m') X(m-m',t)X(m',t)\mathrm{d}m'
\end{aligned} \label{eq:IaXt}
\end{equation}
is a function of the whole size distribution.
We introduce the new variable, 

\begin{equation}
\mathrm{d}\chi= g_\mathrm{local} n_\mathrm{H} \mathrm{d}t.
\end{equation}
We then have
\begin{equation}\label{EqReduced}
  \frac{\mathrm{d}X(a,\chi)}{\mathrm{d}\chi} = C I(a,X,\chi).
\end{equation}

$\chi$ captures the history of the grains, so that it can be used as a sole tracker of the evolution of the distribution. At a constant $\chi$, the state of the distribution is therefore independent of the path taken for a given initial distribution. Equation~(\ref{EqReduced}) then needs to be integrated only once.
This method can be simply included in hydrodynamics simulations by tracking the evolution of the function $\chi,$ and the associated distribution that would have been precalculated can be listed in a table. It is impossible to use it in conjunction with other processes modifying the grain size distribution, however. We detail this caveat in section \ref{SecCaveats}.

\subsection{Application to a turbulent kernel}

The collision kernel derived by \citet{2007A&A...466..413O} assumes that grains are accelerated by the gas turbulence and couple to eddies at different scales, introducing a differential velocity between grains of different sizes. The coagulation kernel reads

\begin{equation} \label{EqKernel}
  K(m,m')=\sqrt{\frac{8}{3\pi}} \sigma \Delta V,
\end{equation}
with $\Delta V$ the differential velocity between the two grains, $\sigma=\pi (a+a')^2$ the collision cross section of the two grains, and $a$ and $a'$ their respective radius. Here we assume that $a>a'$.
We assume the intermediate-coupling regime, in which the stopping time of the grain $\tau_\mathrm{s}$ exceeds the dissipation timescale of the turbulence $\tau_\eta$ , but is shorter than the turbulence-forcing timescale $\tau_\mathrm{L}$. We detail the environments in which this assumption is valid in Sect.~\ref{SecCaveats}. The velocity drift is \citep{2007A&A...466..413O}

\begin{equation}
  \Delta V = \left[\frac{3}{2} c_\mathrm{s}^2 z\left( \frac{\tau_\mathrm{s}'}{\tau_\mathrm{L}} \right) \frac{\tau_\mathrm{s}}{\tau_\mathrm{L}}\right]^{1/2},
\end{equation}
 with $c_\mathrm{s}$ the sound speed and $z(x)$ a function that we assume to be equal to a constant $ z_0=2.97$. Assuming the injection scale of the turbulence is equal to the Jeans length $L_J$, as expected in star-forming regions \citep[e.g.,][]{2016ApJ...824...41I}, the forcing timescale for the turbulent cascade would be \citep{2009A&A...502..845O}

\begin{equation}
  \tau_\mathrm{L} = \frac{L_\mathrm{J}}{c_\mathrm{s}} = \frac{1}{2}\sqrt{\frac{\pi}{G\rho}},
\end{equation}
with $G$ the gravitational constant.
The stopping time of grains is given by \citep{epstein1924}

\begin{equation}\label{EqStoppingtime}
  \tau_\mathrm{s} = \frac{\rho_\mathrm{s}a}{\mu m_\mathrm{H} n_\mathrm{H} v_\mathrm{th}},
\end{equation}
with $\rho_\mathrm{s}$ the bulk density of grains, $\mu$ the average atomic weight, $m_\mathrm{H}$ the mass of the hydrogen atom, $v_\mathrm{th}=(8k_\mathrm{B}T/ \pi\mu m_\mathrm{H})^{1/2}$ the thermal velocity of the grains, $k_\mathrm{B}$ the Boltzmann constant, and $T$ the temperature. 

We define

\begin{equation}
    C_2=\left(\frac{3}{\sqrt{8}} z_0 [k_\mathrm{B}G]^{1/2}
    \frac{\gamma \rho_\mathrm{s}}{\mu m_\mathrm{H}}
  \right)^{\frac{1}{2}},
\end{equation}
with $\gamma$ the adiabatic index of the gas. This is time invariant so long as the grain and gas compositions do not change. After some algebra, we can show that

\begin{equation} \label{Eqdeltv}
  \Delta V = C_2 n_\mathrm{H}^{-\frac{1}{4}} T^{-\frac{1}{4}} a^{\frac{1}{2}},
\end{equation}
or alternatively, assuming $\gamma=5/3$, $\rho_\mathrm{s}=2.3$ g cm$^{-3}$ and $\mu=2.3$,

\begin{equation} 
  \Delta V = (1.74 \times 10^2 \mathrm{cm\ s}^{-1}) \left(\frac{n_\mathrm{H}}{10^4\ \mathrm{cm}^{-3}}\right)^{-\frac{1}{4}} \left(\frac{T}{10\ \mathrm{K}}\right)^{-\frac{1}{4}} \left(\frac{a}{10\ \mathrm{nm}}\right)^{\frac{1}{2}}.
\end{equation}
As assumed in section \ref{SecGeneralform}, the coagulation kernel can thus be written

\begin{equation}
  K(m,m') = C g_\mathrm{local}(n_\mathrm{H},T) h(a,a'),
\end{equation}
where
    \begin{align}
      g_\mathrm{local}(n_\mathrm{H},T) &= n_\mathrm{H}^{-\frac{1}{4}} T^{-\frac{1}{4}},\\
      h(a,a') &= (a + a')^2a^{1/2},\\
      C &= (8 \pi/3)^{1/2} C_2.
    \end{align}
We therefore have
\begin{equation}
  \frac{\mathrm{d}X(a,\chi)}{\mathrm{d}\chi} = C I(a,X,\chi),
\end{equation}
with
\begin{equation}
\mathrm{d}\chi= n_\mathrm{H}^{\frac{3}{4}} T^{-\frac{1}{4}} \mathrm{d}t.
\end{equation}

\subsection{Tests}\label{SecTestcoag}

\subsubsection{Methods}

To test the validity of our derivation, we compared the state of a size distribution at a given $\chi$ in different contexts. We used our new code Ishinisan, a coagulation algorithm inspired by the Dustdap code \citep{2007A&A...476..263G}, that we make public with the publication of this paper. We emphasize that Ishinisan computes the (discrete version of) Smoluchowski equation (\ref{EqSmolu}) and not the derived equation (\ref{EqReduced}). The code is able to reproduce the coagulation of grains using the constant and additive kernels for which analytical solutions exist. The aim of these tests is to show the validity of our derivation that the coagulated size distribution only depends on the value of $\chi$. Appendix \ref{AppIshinisan} details the methods we used that are implemented in the code.

We took an MRN distribution with $a_\mathrm{min}=5$ nm and $a_\mathrm{max}=250$ nm as our initial condition. The number density of grains is determined by the slope of the distribution

\begin{equation}
  \frac{dn}{da} = K a^{-3.5},
\end{equation}
with $K$ a constant that is constrained by the total mass of grains (the dust-to-gas ratio $d$), and the minimum and maximum sizes of the distribution (see Appendix \ref{AppIshinisan} for the calculation). We took the usual value $d=0.01$, and we again assumed $\rho_\mathrm{s}=2.3$ g cm$^{-3}$, $\mu=2.3,$ and $\gamma=5/3$.

\subsubsection{Comparing various environments}

Our reference environment was taken from a numerical simulation of a protostellar collapse using the RAMSES code \citep{Teyssier}. We tracked the density and temperature history of a tracer particle during the first phases of the collapse of a uniform $M=1$ M$_\odot$ sphere of gas until its arrival at the first \citet{Larson1969} core. The trajectory is mostly isothermal at $T=10$ K, with the temperature rising as the density reaches $n_\mathrm{H} \approx 2\times 10^{11}$ cm$^{-3}$. The initial density was $n_\mathrm{H} = 1.4 \times 10^6$ cm$^{-3}$ , and we stopped the simulation when the particle was well settled into the first core at $n_\mathrm{H}=2.4 \times 10^{12}$ cm$^{-3}$. The integration along the path gives $\chi=8.12\times 10^{17}$ cgs.
We used Ishinisan to compute the coagulation along the trajectory of the particle and to compare it with environments at constant temperature and density (physically realistic or not) that are summarized in Table \ref{TableEnvironments}. The time in each case was chosen to yield the same value of $\chi$.

\begin{figure*}
\begin{center}
\includegraphics[trim=3cm 1cm 3cm 1.5cm, width=0.9\textwidth]{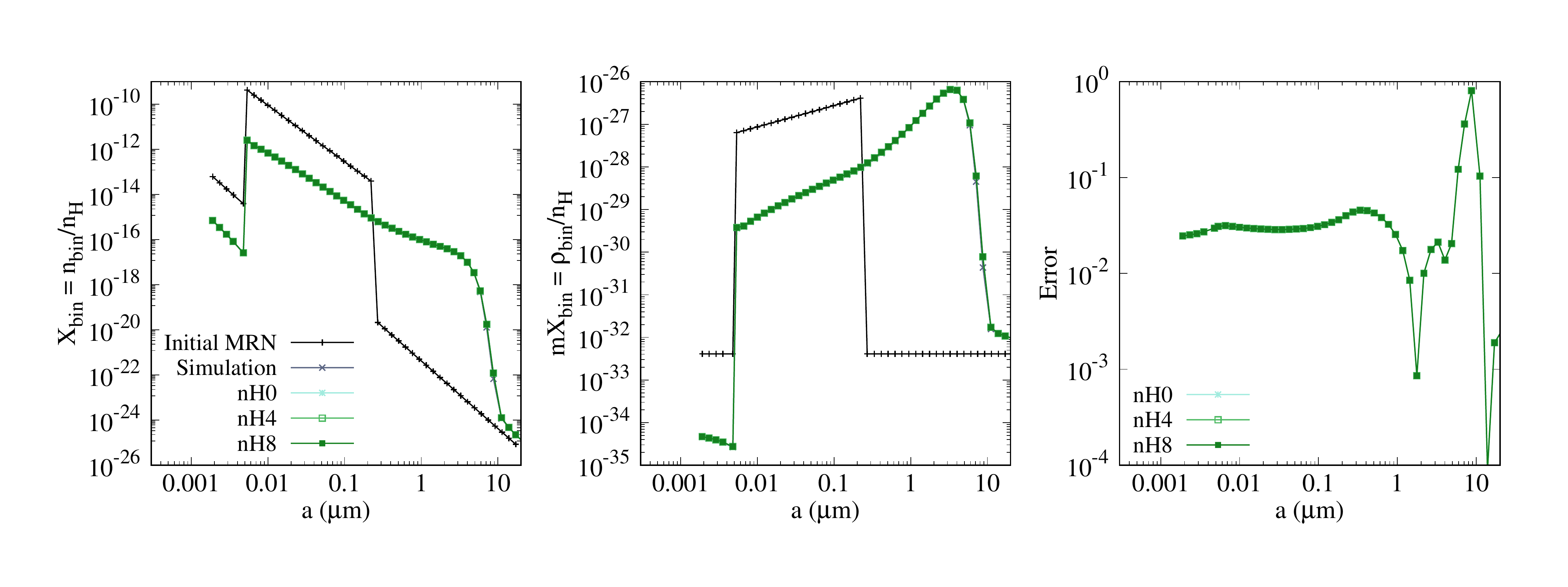}
  \caption{Fractional abundances $X(a)=n(a)/\nh$ (left panel) and fractional mass $\rho(a)/\nh$ of grain bins for the cases listed in Table \ref{TableEnvironments}. The black line represents the initial (MRN) distribution, while the other lines give the distribution at $\chi=8.12 \times 10^{17}$ cgs. The right panel shows the relative error of the constant density cases nH0, nH4, and nH8 (indistinguishable to machine precision) compared to the simulation Sim.}
  \label{FigDistribtest}
\end{center}
\end{figure*}

\begin{table}
  \caption{Test coagulation environments}
\label{TableEnvironments}
\centering
\begin{tabular}{llll}
\hline\hline
  Case & $\nh$ (cm$^{-3}$) & T (K) & Time (yr) \\
\hline
  Sim  &  $1.4 \times 10^6$ - $2.4\times 10^{12}$ & $10$ - $70$ & $3.26 \times 10^4$  \\
  nH0  &  $10^0$                                  & $10^5$      & $4.58 \times 10^{11}$ \\
  nH4  &  $10^4$                                  & $10$        & $4.58 \times 10^7$  \\
  nH8  &  $10^8$                                  & $10 $       & $4.58 \times 10^4$  \\
\hline
\end{tabular} 
\end{table}

The size distributions at $\chi=8.12 \times 10^{17}$ cgs are represented in Figure \ref{FigDistribtest} in number density and mass density (left and middle panel, respectively), with the error displayed in the right panel. The three cases at constant density are indistinguishable and only show a difference of a few percent from the virtual particle of the simulation. Around $a=10$ $\micro$m, the error almost reaches 100\% for one of the bins. The discrepancies are due to a minor difference between the equivalent $\chi$ of the simulation and the theoretical cases. The error is also exacerbated by the steepness of the distribution at this location and the low density of the bin. In the MRN and in the coagulated distribution, the small grains dominate in number, but the large grains dominate in mass.

\section{Grain ionization}\label{SecIonization}

We now determine the ion and electron number density, $\ni$ and $\ne$, and the charge $Z_k$ of each grain size, knowing the grain size distribution, the CR ionization rate $\zeta$, the mean atomic mass of ions $\mu_\mathrm{i}$, and the gas density $\nh$ and temperature $T$. We took $a_k$ to be the radius of grains in size bin $k$ and $n_k$ their number density. \citet[][hereafter DS87]{DraineSutin} introduced the reduced temperature of grains, 

\begin{equation}
  \tau_k = \frac{a_k k_\mathrm{B} T}{e^2},
\end{equation}
with $e$ the elemental electric charge. DS87 distinguished two regimes: $\tau_k \gg 1$ for high temperature and large grains, and $\tau_k \ll 1$ for low temperature and small grains. They calculated the equilibrium charge distribution of grains $f$ by equating the flux of ions $J_\mathrm{i}$ and electrons $J_\mathrm{e}$ on grains. For a given grain of temperature $\tau$ with charge $Z$, we have

\begin{equation}\label{EqFluxequal}
  f(Z)J_\mathrm{i}(\tau_k,Z) = f(Z+1)J_\mathrm{e}(\tau_k,Z+1),
\end{equation}
where

\begin{equation}
  J_s(\tau_k,Z) = n_s s_s \sqrt{\frac{8k_\mathrm{B}T}{\pi m_s}}\pi a^2 \tilde{J}(\tau_k, Ze/q_s).
\end{equation}
The subscript $s$ stands for either ions or electrons, $s_s$ is the sticking coefficient of the species $s$ onto grains, and $m_s$ is the mass per particle of species $s$ (so $m_i = \mu_i m_\mathrm{H}$). The polarization enhancement factor $\tilde{J}$ is given by DS87 Equations [3.3], [3.4], and [3.5] and is a function of the ratio of the charge of the grain $Ze$ and the charge of the particle $q_s$. For an attractive polarization, $Ze/q_s<0$. The system is closed by the normalization 

\begin{equation}
  \sum_Z f(Z) = 1.
\end{equation}
In the following, we assume $s_\mathrm{i}=1$, and we define
\begin{equation}
\Theta=s_\mathrm{e}(\mu_\mathrm{i}m_\mathrm{H}/m_\mathrm{e})^{1/2}.
\end{equation}

We assumed that the CRs are the exclusive ionization source. The ionization rate equals the recombination rate of ions in the gas phase (with electrons) and onto (negatively charged) grains,

\begin{equation} \label{EqRecombination}
  \zeta \nh = \svie \ni \ne + \ni v_\mathrm{i} \sum_k \langle\tilde{J}(\tau_k)\rangle n_k \pi a_k^2,
\end{equation}
where $\svie = 2\times 10^{-7} [T/300]^{-1/2}$ is the collision rate between ions and electrons, $v_\mathrm{i}=[8k_\mathrm{B}T/(\pi\mu_\mathrm{i}m_\mathrm{H})]^{1/2}$ is the thermal velocity of ions, and $\langle\tilde{J}(\tau_k)\rangle$ is the average polarization factor for a grain of temperature $\tau_k$. Finally, the charge neutrality condition imposes

\begin{equation}\label{EqElectroneutral}
  \ni - \ne + \sum_k n_k Z_k = 0.
\end{equation}

\subsection{Large grains and high temperatures $\tau_k \gg 1$} \label{SecTauhigh}

For high temperatures and large grains, DS87 showed that the charge distribution is a Gaussian, whose average value is

\begin{equation}\label{EqZktauhigh}
  Z_k = \psi \tau_k < 0,
\end{equation}
where $\psi$ represents the ratio between the electric potential of the grain and the kinetic energy of electrons. It is the solution to the equation \citep[][DS87]{1941ApJ....93..369S} 
\begin{equation} \label{EqPsi}
  1-\psi = \Theta \frac{\ne}{\ni} \mathrm{e}^{\psi}.
\end{equation}

The combination of Equation (\ref{EqZktauhigh}) with Equation (\ref{EqElectroneutral}) for charge neutrality gives
\begin{equation} \label{EqPsitauhigh}
  \ni-\ne=-\psi\sum_k n_k \tau_k.
\end{equation}

The average enhancement factor is given by (DS87)

\begin{equation}
 \langle\tilde{J}(\tau_k \gg 1)\rangle = 1-\psi.
\end{equation}
If we define $\epsilon = \ne / \ni <1$, then the $\psi$ Equation (\ref{EqPsi}), the charge neutrality Equation (\ref{EqPsitauhigh}) and the recombination Equation (\ref{EqRecombination}) become the following system of four equations and four unknowns:

\begin{equation}
\begin{aligned}
  \epsilon =& \frac{1-\psi}{\Theta\mathrm{e}^{\psi}}, \\
  \ni =& -\psi \frac{\sum_k n_k \tau_k}{1-\epsilon},\\
  \ne =& \epsilon \ni,\\
  1 =& \frac{\svie \epsilon \ni^2}{\zeta \nh} + (1-\psi) \frac{\ni v_\mathrm{i} \sum_k n_k \pi a_k^2}{\zeta \nh},
\end{aligned}
\end{equation}
which is equivalent to a unique equation in $\psi$ that can be numerically solved.

\subsection{Small grains and low temperatures $\tau_k \ll 1$} \label{SecTaulow}

Small grains can only carry one electric charge due to their strong polarization. Equation (\ref{EqFluxequal}) can therefore be written

\begin{align}
  f(-1) &= f(0) \frac{J_\mathrm{e}(0)}{J_\mathrm{i}(-1)} = f(0) \epsilon \Theta\frac{\tilde{J}(\tau_k,0)}{\tilde{J}(\tau_k,-1)}, \label{Eqfm1f0}\\
  f(1) &= f(0) \frac{J_\mathrm{i}(0)}{J_\mathrm{e}(1)} = f(0) \frac{1}{\epsilon \Theta}\frac{\tilde{J}(\tau_k,0)}{\tilde{J}(\tau_k,-1)}, \\
  f(-1) &+ f(0) + f(1) =1.
\end{align}
For $\tau_k \ll 1$, the polarization factors can be approximated to
\begin{equation}
  \tilde{J}(\tau_k,0)  \approx \left( \frac{\pi}{2\tau_k}
                        \right)^{\frac{1}{2}}, \mbox{ and }
  \tilde{J}(\tau_k,-1)  \approx \frac{2}{\tau_k}.
\end{equation}

DS87 assumed a regime in which $\ne \approx \ni$, and therefore $f(1) \ll f(-1)$. However, this is not true at high density because the recombination rate increases faster than the ionization rate, and the grains absorb most of the electrons. The number of grains increases with density, and their average charge plummets toward zero. Therefore we cannot neglect the positively charged grains here and we deviate from the solutions derived by DS87. \citet{2016ApJ...833...92I} investigated this regime by calculating the deviation of the charge distribution compared to a pure ion-electron plasma. They did not include the grain polarization in their study, however. \citet{1993A&A...280..617O} also performed a similar calculation, accounting for the polarization, to determine the impact of the grain charges on their coagulation rate. They found that this factor had only a limited effect on the collision probability of grains, however.

The above system can be solved into

\begin{align}
  f(0) &=  \frac{1}{ 1 + \frac{1}{\alpha_k}\left[\epsilon \Theta +\frac{1}{\epsilon \Theta} \right]},\\
  f(-1) &= \frac{1}{ 1 + \frac{1}{\epsilon \Theta} \alpha_k + \frac{1}{\epsilon^2 \Theta^2}},\\
  f(1) &= \frac{1}{ 1 + \epsilon \Theta \alpha_k + \epsilon^2 \Theta^2},
\end{align}
where $\alpha_k = [8/(\pi \tau_k)]^{1/2}$. 
The average charge of grains is then just given by
\begin{equation}\label{EqZktaulow}
\begin{aligned}
  Z_k &= \sum_Z Z f(Z) = f(1) - f(-1) \\
      &= \frac{1-\epsilon^2 \Theta^2}{1+ \epsilon \Theta \alpha_k + \epsilon^2 \Theta^2}.
\end{aligned}
\end{equation}
The average polarization factor is calculated similarly. For the recombination of ions onto grains,

\begin{equation}\label{EqJktaulow}
  \langle\tilde{J}(\tau_k)\rangle = \sum_Z \tilde{J}(\tau_k, Ze/q_\mathrm{i}) f(Z) = \tilde{J}(\tau_k,-1)f(-1)+\tilde{J}(\tau_k,0)f(0).
\end{equation}
We can neglect $\tilde{J}(\tau_k,1)f(1)$ because it represents the repulsive interaction between a positive ion and a positively charged grain. Combining Equations (\ref{EqJktaulow}) and (\ref{Eqfm1f0}) gives

\begin{equation}\label{EqJktaulow2}
\begin{aligned}
  \langle\tilde{J}(\tau_k)\rangle &= \tilde{J}(\tau_k,-1)f(-1) \left[ 1 + \frac{1}{\epsilon \Theta} \right],\\
                                  &= \frac{\frac{2}{\tau_k}\left[ 1 + \frac{1}{\epsilon \Theta} \right]} {1 + \frac{1}{\epsilon \Theta}\alpha_k + \frac{1}{\epsilon^2 \Theta^2}}.
\end{aligned}
\end{equation}
Neglecting $f(1)$ and $\tilde{J}(\tau_k,0)f(0)$ in this calculation gives the results derived by DS87 (their Eqs.\ [4.12] and [4.13] for $f(-1)$ and $f(0)$, and Eqs.\ [5.3] and [5.11] for $Z_k$ and $\langle \tilde{J}(\tau_k) \rangle$).

\subsection{General case} \label{SecTauall}

DS87 proposed a generalization for all $\tau$ of the electric charge and the enhancement factor, which is simply the sum of both cases (Eqs.\ [\ref{EqZktauhigh}] and [\ref{EqZktaulow}]):

\begin{equation}\label{EqZktauall}
Z_k = \psi \tau_k + \frac{1-\epsilon^2 \Theta^2}{1+ \epsilon \Theta \alpha_k + \epsilon^2 \Theta^2}.
\end{equation}
Similarly, the recombination enhancement factor is

\begin{equation}\label{EqJktauall}
  \langle \tilde{J}(\tau_k) \rangle = (1-\psi) + \frac{\frac{2}{\tau_k}\left[ 1 + \frac{1}{\epsilon \Theta} \right]} {1 + \frac{1}{\epsilon \Theta}\alpha_k + \frac{1}{\epsilon^2 \Theta^2}}.
\end{equation}
Combining these two expressions with Equation (\ref{EqPsi}) for $\psi$, Equation (\ref{EqElectroneutral}) for charge neutrality and Equation~(\ref{EqRecombination}) for ionization gives the following system in $\psi$, $\epsilon,$ and $\ni$:

\begin{equation}\label{EqIonizgeneral}
  \begin{aligned}
    \epsilon =& \frac{1-\psi}{\Theta\mathrm{e}^{\psi}}, \\
    \ni=&  -\frac{1}{1-\epsilon} \sum_k n_k Z_k,\\
    1=& \frac{\svie \epsilon \ni^2}{\zeta \nh} + \frac{\ni v_\mathrm{i}}{\zeta \nh} \sum_k n_k \pi a_k^2\langle\tilde{J}(\tau_k)\rangle,
  \end{aligned}
\end{equation}
which is equivalent to a unique equation in $\psi$. This system can be numerically integrated to obtain $\psi$, then $\ni$, $\epsilon$, $\ne$ , and all $Z_k$. We describe a method for a solution in Appendix \ref{AppNumerical}.

\subsection{Test}\label{SecTestioniz}

In this section, we compare the ionization given by the system of equations in the general case (Eq.\ \ref{EqIonizgeneral}) to the results of the Dustdap code \citep{2007A&A...476..263G,2020A&A...643A..17G}, which has been developed to study the detailed charge and dynamics of a full size distribution of dust. 
We set up an MRN distribution between $a_\mathrm{min}=5$ nm and $a_\mathrm{max}=250$ nm with a dust-to-gas mass ratio of $d=0.006$. The size distribution was divided into 50 bins of equal width in log space and did not evolve for this test. For densities between $\nh = 10^4$ cm$^{-3}$ and $\nh = 10^{12}$ cm$^{-3}$, we calculated the average electric charge of each bin of grain size as well as $\ni$ and $\ne$. We assumed a CR ionization rate of $\zeta=5\times 10^{-17}$ s$^{-1}$, an average ion mass of $\mu_\mathrm{i}=25,$ and $s_\mathrm{e}=0.5$. The grains have an ice mantle of thickness $8.7$ nm and negligible mass, which does not affect the calculation except for the grain radii. We used the Newton-Raphson algorithm described in Appendix \ref{AppNumerical}.

\begin{figure}
\begin{center}
\includegraphics[trim=2.5cm 0cm 1cm 1.5cm, width=0.45\textwidth]{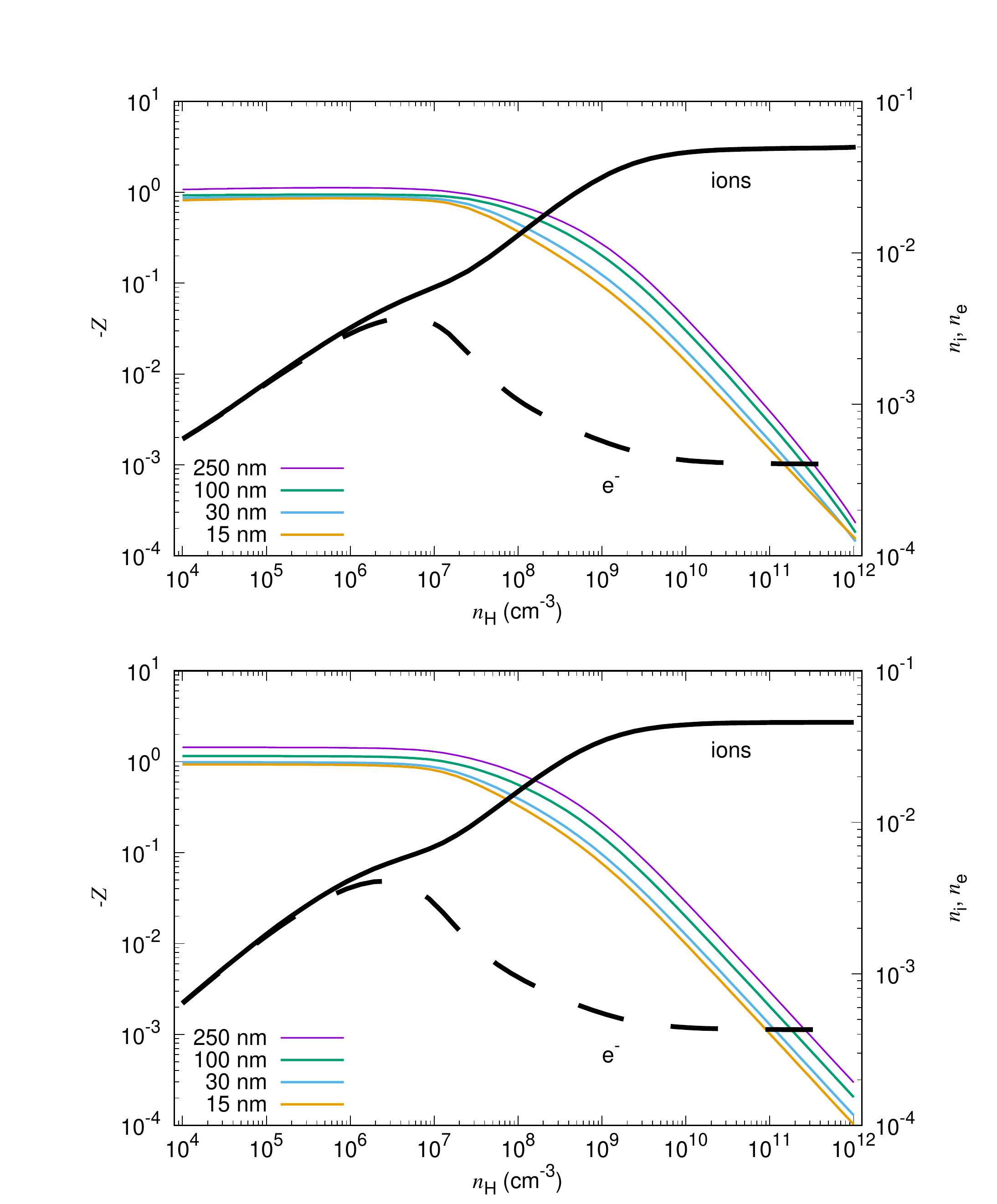}
  \caption{Evolution of the average charge of grains (colored lines, left axis) and the density of ions and electrons (solid and dashed black lines, right axis) as a function of density. Comparison between Dustdap (top panel) and the method presented in this work (bottom panel).}
  \label{FigIontest}
\end{center}
\end{figure}

The results are displayed in Figure \ref{FigIontest}. Our method (bottom panel) shows an excellent agreement with Dustdap (top panel). Initially, the average charge of the grains (which are small, $\tau < 0.15$) is $\approx -1,$ and electrons and ions are present in equivalent numbers. As density increases, the recombination becomes much more efficient than the ionization, and the average charge of the grains decreases because their number increases. Most of the electrons are captured by the grains, and their fractional abundance decreases. At high density, the ratio $\ni/\ne$ converges toward $\Theta \approx 107$.

\section{Caveats}\label{SecCaveats}

\subsection{Intermediate coupling assumption}

This coagulation method assumes that the grains are in the intermediate coupling regime with the turbulence. This hypothesis is valid only if the grain stopping time is longer than the dissipation timescale of the turbulence $\tau_\eta$, but shorter than the turbulence injection timescale $\tau_\mathrm{L}$. The dissipation timescale depends on the Reynolds number as
\begin{align}
  \tau_\eta =& \frac{\tau_\mathrm{L}}{\sqrt{\mathrm{Re}}}, \\
  \mathrm{Re} =& 6.2 \times 10^7 \left(\frac{n_\mathrm{H}}{10^5
                 \mbox{ cm} ^{-3}}\right)^{1/2}.
\end{align}

\begin{figure}
\begin{center}
\includegraphics[trim=1.5cm 0.5cm 1cm 1cm, width=0.49\textwidth]{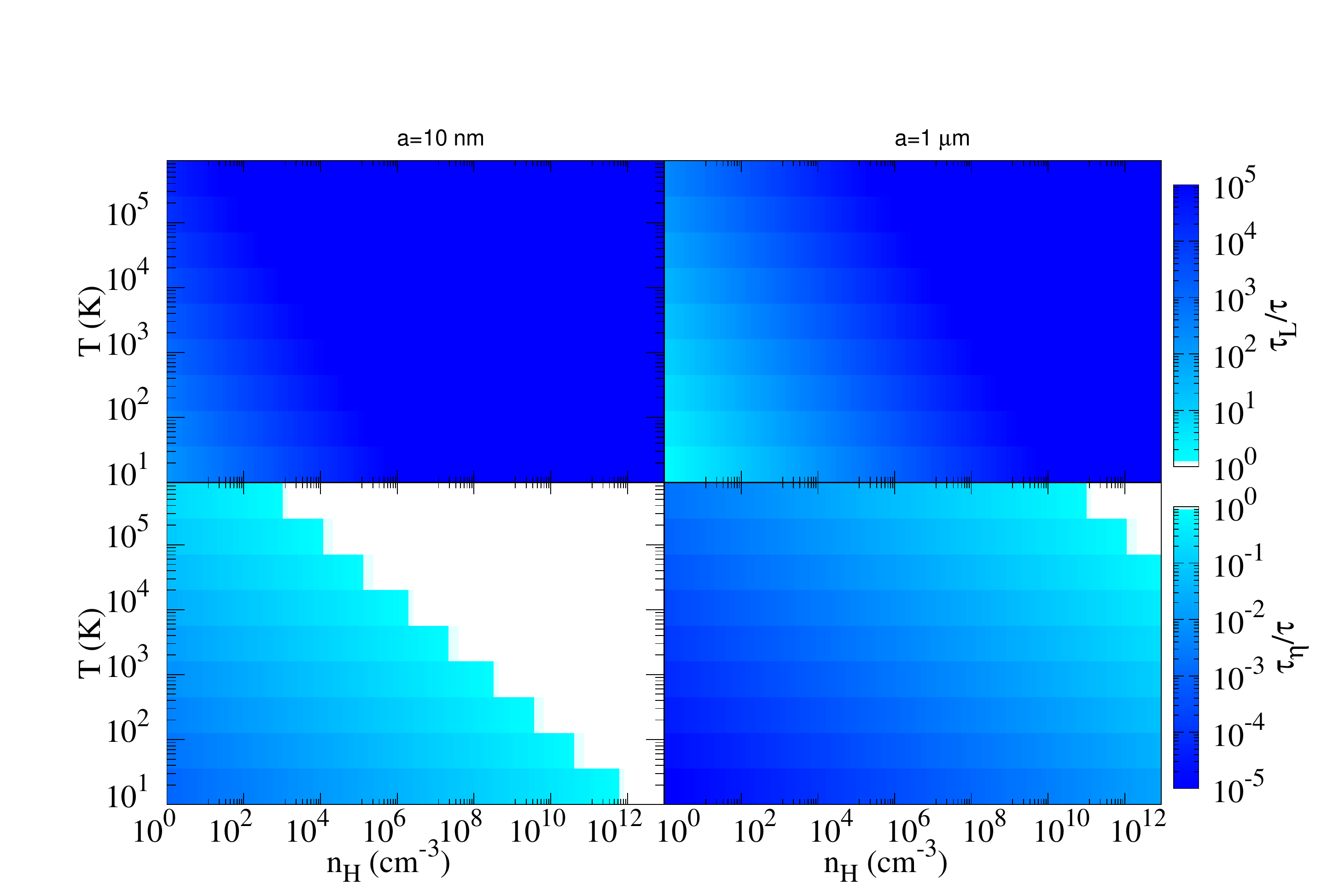}
  \caption{Ratio $\tau_\mathrm{L} / \tau$ (top) and $\tau_\eta / \tau$
    (bottom) in density-temperature intervals for grain sizes of
    $a=10$ nm (left) and $a=1$ $\micro$m (right). Blue
    represents the domain of validity for the intermediate coupling assumption.}
  \label{FigValid}
\end{center}
\end{figure}

The condition $\tau_\eta < \tau_\mathrm{s}< \tau_\mathrm{L}$ constrains the validity domain of the intermediate coupling hypothesis, which depends on the density, temperature, and grain size. Figure \ref{FigValid} shows the ratio $\tau_\mathrm{L}/\tau_\mathrm{s}$ (top) and $\tau_\eta/\tau_\mathrm{s}$ (bottom) in the $n_\mathrm{H}$-$T$ domain for $a=10$ nm (left) and $a=1$ $\micro$m (right). Blue means that the condition is satisfied, while white shows where the hypothesis is not valid (in these cases, the grains would fall into the tightly coupled regime). The only problematic domains are in the high density-high temperature quadrants, especially for the smaller grains. However, this domain corresponds to protostellar or protoplanetary disk conditions. The grains would have experienced significant coagulation by the time they reach the protostar, as shown by Figure \ref{FigDistribtest}. This point will be developed further in a forthcoming paper.

\subsection{Fragmentation and other processes}

The $\chi$ variable does not account for and cannot be used with other processes that dust may undergo in addition to coagulation, in particular, it cannot be used with fragmentation. Grains also grow by accreting gaseous material from their surroundings, especially during their long journey through the ISM \citep{2018ApJ...857...94Z}. Accretion severely depletes the number of very small grains as the radius growth rate is independent of the grain size. On the other hand, sputtering \citep{1989IAUS..135..431M,2016P&SS..133..107M} is a process that decreases the grain radius, especially in supernova shocks in the ISM. In more dense environments and at low temperature, such as in protostellar cores, grains accrete ice mantles \citep{2015A&A...576A..49H}. The larger resulting cross section increases the collision rate, but the mantle does not necessarily facilitate the sticking of grains during a collision \citep{2020arXiv200805841K}.

We considered purely spherical grains for simplicity. However, coagulation tends to produce fractal structures \citep{2009A&A...502..845O}, changing the effective cross section and density of the grain. These issues can be mitigated by introducing a porosity factor, although nonspherical shapes also promote fragmentation.

\begin{figure}
\begin{center}
\includegraphics[trim=3cm 5.5cm 1cm 3cm, width=0.49\textwidth]{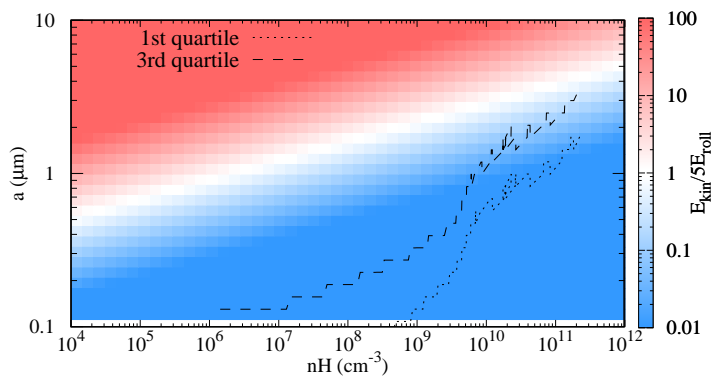}
  \caption{Ratio $E_\mathrm{kin}/5E_\mathrm{roll}$ as a function of density (x-axis) and the largest of the two colliding grains (y-axis). Red indicates a ratio above 1, meaning that the collision restructures or fragments the grains. The dotted and dashed lines are the first and third quartile of the MRN size distribution that coagulate during the protostellar collapse presented in section \ref{SecTestcoag}.}
  \label{Figfragment}
\end{center}
\end{figure}

\citet{2009A&A...502..845O} provided conditions for the fragmentation of aggregates based on \citet{1997ApJ...480..647D}. If the kinetic energy of the collision $E_\mathrm{kin}$ exceeds five times the rolling energy $E_\mathrm{roll}$ of the grain, then the aggregate experiences significant structure changes or breaks.
The kinetic and rolling energies are given by
\begin{align}
  E_\mathrm{kin} &= \frac{1}{2}\frac{mm'}{m+m'}\Delta v^2,\\
  E_\mathrm{roll} &= 6 \pi^2 \xi_\mathrm{crit} \gamma_s a_\mu, 
\end{align}
where $m$ and $m'$ are the masses of the two colliding grains and $\Delta v$ their collision velocity (given by Equation [\ref{Eqdeltv}], which depends only on the size of the largest grain). The critical displacement for irreversible rolling $\xi_\mathrm{crit}=2\times 10^{-7}$ cm, $\gamma_s$ the surface energy density of the material, and $a_\mu$ the reduced radius of the grain monomers. We took $\gamma_s=370$ erg s$^{-1}$ for ice-coated silicates and assumed $T=10$ K. As was done by \citet{2009A&A...502..845O}, we considered $0.1~\micro$m monomers, therefore $a_\mu=0.05~\micro$m. Figure \ref{Figfragment} shows the ratio $E_\mathrm{kin}/5E_\mathrm{roll}$ as a function of density and the largest grain size involved in the collision. In the blue region, the ratio is lower than 1 and the energy of the collision is not high enough to restructure the grain, while the red region is the opposite. We superimposed the first and third quartile of the MRN size-distribution presented in the test of Sect \ref{SecTestcoag} that coagulates during the protostellar collapse \footnote{That is, 25\% of the mass is below the dotted line, while 75\% is below the dashed line}.

At the densities of protostellar envelope ($\nh < 10^{10} \cc$), fragmentation seems to be a minor process even with grain growth. However, at higher densities ($\nh > 10^{10} \cc$), typical of the first Larson core and the disk, the largest grains of the coagulated distribution reach the threshold. Therefore we cannot rule out fragmentation in the disk, even at early phases, which is also supported by observations of small grains in protoplanetary disks \citep{2004A&A...427..179H,2006Sci...314..621L}. For bare silicates (without ice), the fragmentation threshold is lower but the conclusion is the same.
This process tends to steepen the distribution as larger grains shatter into smaller grains. While the mass remains in the large grains, the small grains dominate the number density and the surface area, hence significantly impacting the non-ideal MHD resistivities and the opacity of the medium. The results given by the method presented in this paper should thus be taken with caution in the environment of a protostellar disk.

We did not consider differential dynamics of gas and dust either. This can generate local variations of the dust-to-gas ratio and the dust distribution for grains larger than $\approx 50~\micro\meter$ during protostellar collapse \citep{2017MNRAS.465.1089B,2019A&A...626A..96L,2020A&A...641A.112L} and $\approx~10 \micro \meter$ in molecular clouds \citep{2017MNRAS.471L..52T}.

\subsection{Limitations of the ionization model}

The ionization formula derived by DS87 does not account for charge transfer between grains. As \citet{2016A&A...592A..18M} pointed out in their Appendix B, this would lead to inaccurate values of the ionization and the MHD resistivities if the grains were the dominant charge carrier. This could in principle be the case during the protostellar collapse, at the end of the isothermal phase, and during the life of the first Larson core \citep{Larson1969}. However, as \citet{2020A&A...643A..17G} showed, coagulation efficiently reduces the number of grains, ensuring that ions remain the dominant charge carriers in the gas during collapse and in the protoplanetary disk \citep[see also][]{Okuzumi2009,Fujii2011}.

Finally, our ionization model assumes that CRs are the only source of ionization. The system of equations we derive needs to be adapted for other ionization sources, such as UV or X-rays from young stars.

\section{Conclusion}\label{SecConclusion}

We presented two analytical methods.

With the first method, we have derived a new reduced variable to track grain coagulation for any kernel in which the variable can be separated. Given an initial size distribution, the coagulated distribution depends only on the value of an integral $\chi$ along the trajectory of the grains. We applied this method to the turbulent kernel derived by \citet{2007A&A...466..413O} and found $\chi = \int_0^t n_\mathrm{H}^{3/4} T^{-1/4} dt$. Although other mechanisms that process grains, such as fragmentation, cannot be included in the calculation, the $\chi$ variable provides a simple and inexpensive technique to evaluate the role of grain coagulation in analytical and numerical studies.

We also described a method for computing the ionization of a medium knowing the ionization rate, the gas density and temperature, and the grain size distribution. We adapted the model of DS87 by reducing it to a single equation that can be numerically solved to rapidly compute the average charge of grains as a function of their size, as well as the ion and electron density. Using a Newton-Raphson scheme, the solution of this equation is fast enough to be performed on the fly in numerical simulations.

Two upcoming papers will focus on the application of these methods to the formation of dense molecular cores and in protostellar collapse simulations. These tools can be used together to rapidly and self-consistently calculate the nonideal MHD resistivities locally.
The code Ishinisan is publicly available at https://bitbucket.org/pmarchan/ishinisan.

\begin{acknowledgements}
P. M. acknowledges financial support by the Kathryn
W. Davis Postdoctoral Fellowship of the American Museum of Natural
History. U. L. acknowledges financial support from the
European Research Council (ERC) via the ERC Synergy Grant ECOGAL
(grant 855130). M.-M. M. L. acknowledges partial support from NSF
grant AST18-15461 and NASA Theory Program grant NNX17AH80G.   
\end{acknowledgements}

\appendix

\section{Numerical solution for ionization}\label{AppNumerical}

We describe a numerical solution to find the values of $\psi$, $\ni$ and $\ne$ of the system of equations (\ref{EqIonizgeneral}).
These can be rewritten as
\begin{align}
  Z_k =& \psi \tau_k + \frac{1-\epsilon^2 \Theta^2}{1+ \epsilon \Theta \alpha_k + \epsilon^2 \Theta^2},\label{EqsysZ}\\
\langle \tilde{J}(\tau_k) \rangle =& (1-\psi) + \frac{\frac{2}{\tau_k}\left[ \epsilon^2\Theta^2 + {\epsilon \Theta} \right]} {1 + \epsilon \Theta\alpha_k + \epsilon^2 \Theta^2},\label{EqsysJ}\\
\epsilon =& \frac{1-\psi}{\Theta\mathrm{e}^{\psi}}, \label{EqsysEps}\\
\ni=&  -\frac{1}{1-\epsilon} \sum_k n_k Z_k,\label{EqsysNi}\\
f(\psi)=& \frac{\svie \epsilon \ni^2}{\zeta \nh} + \frac{\ni v_\mathrm{i}}{\zeta \nh} \sum_k n_k \pi a_k^2\langle\tilde{J}(\tau_k)\rangle -1 = 0,\label{EqsysF}
\end{align}

The two unknowns $\epsilon$ and $\ni$ are functions of $\psi$, therefore we need to solve $f(\psi)=0$. The charge distribution $f$ is monotonous on its validity interval $[\psi_0,0[$, where $\psi_0$ is the solution of Equation (\ref{EqPsi}) for $\epsilon=1$ ($\psi_0=-2.504$ for $\mu_\mathrm{i}=1$ and $\psi_0=-3.799$ for $\mu_\mathrm{i}=25,$ e.g.).

We used a Newton-Raphson iteration scheme to solve the equations. Starting from an initial  value of $\psi =\psi_0$ or the last calculated $\psi$, we calculated the next estimate
\begin{equation}\label{EqAppNR}
  \psi_{n+1} = \psi_n - \left( \frac{f(\psi_n)}{\mathrm{d}f/\mathrm{d}\psi(\psi_n)}\right),
\end{equation}
and we iterated until $|f(\psi_n)| < \delta$, where $\delta \ll 1$ is a
chosen tolerance factor.

The first derivative of the function $f(\psi)$ reads
\begin{equation}
\begin{aligned}
\label{EqAppDfdpsi}
\frac{\mathrm{d}f}{\mathrm{d}\psi}(\psi) = & \frac{\svie \ni}{\zeta \nh} \left( \ni \depsdpsi + 2\epsilon \dnidpsi \right)\\
&+ 2\frac{\ni v_\mathrm{i}}{\zeta \nh}\left(\epsilon^2 \Theta^2 + \epsilon \Theta\right)\left[ \left( \frac{1}{\ni}\dnidpsi + \frac{2\epsilon \Theta +1}{\epsilon(1+\epsilon \Theta)}\depsdpsi\right)\right.\\
&\sum_k \frac{n_k \pi a_k^2}{\tau_k \left(1 + \epsilon \Theta \alpha_k + \epsilon^2 \Theta^2 \right)} \\
&\left. + \depsdpsi \sum_k \frac{n_k \pi a_k^2 \left(2\epsilon \Theta^2 + \Theta\alpha_k\right)}{\tau_k \left(1 + \epsilon \Theta\alpha_k + \epsilon^2 \Theta^2 \right)^2} \right]\\
& +\frac{\ni v_\mathrm{i}}{\zeta \nh} \left( \frac{1}{\ni}\dnidpsi (1-\psi) -1 \right) \sum_k n_k \pi a_k^2  ,
\end{aligned}
\end{equation}
where
\begin{align}
  \depsdpsi &= -\epsilon\frac{2-\psi}{1-\psi}, \label{EqAppdepsdpsi}\\
  \dnidpsi &= -\frac{1}{(1-\epsilon)^2}\depsdpsi \nonumber\\
  & \sum_k \frac{n_k[ -\epsilon^4 \Theta^4 + (4\Theta^2-\Theta^3\alpha_k)\epsilon^2 + (2\Theta\alpha_k -4\Theta^2)\epsilon + 1 -\Theta\alpha_k]}{(1+\epsilon \Theta \alpha_k + \epsilon^2 \Theta^2)^2} \nonumber\\
  & -\frac{1}{1-\epsilon} \left( \frac{\psi}{1-\epsilon}\depsdpsi +1 \right) \sum_k n_k \tau_k.\label{EqAppdnidpsi}
  \end{align}

  The solution algorithm is therefore the following:
\begin{enumerate}
  \item Calculate the quantities $\tau_k$ and $\alpha_k$ for all grains.
  \item Calculate $\psi_0$, the solution of Equation (\ref{EqPsi}) with $\epsilon=1$.
  \item Using $\psi_0$ as a first estimate, calculate in turn
    $\epsilon$ (Eq.\ \ref{EqsysEps}), $Z_k$ (Eq.\ \ref{EqsysZ}), $\ni$
    (Eq.\ \ref{EqsysNi}), $f(\psi)$ (Eq.\ \ref{EqsysF}) , $\mathrm{d}\epsilon/\mathrm{d}\psi$
    (Eq.\ \ref{EqAppdepsdpsi}), $\mathrm{d}n_i/\mathrm{d}\psi$ (Eq.\ \ref{EqAppdnidpsi})
,    and $\mathrm{d}f/\mathrm{d}\psi$ (Eq.\ \ref{EqAppDfdpsi}).
  \item Calculate the new estimate for $\psi$ with Equation (\ref{EqAppNR}).
  \item Repeat steps 3 and 4 with the new value of $\psi$ until $f(\psi)<\delta$.
\end{enumerate}

\section{Methods of Ishinisan}\label{AppIshinisan}

The resolution methods implemented in Ishinisan are largely inspired by the Dustdap code \citep{2007A&A...476..263G,2020A&A...643A..17G}. Ishinisan takes a size distribution as initial condition and computes the evolution of the dust mass-density with time according to the Smoluchowski equation for mass density (equation \ref{EqSmolu}) \footnote{not the derived reduced equation (\ref{EqReduced}) so as to improve the versatility of the code and facilitate mass conservation during the calculations.}. The gas density, temperature, and magnetic field are either fixed or can be input as a datafile tracing its history (from a tracer particle in a simulation, e.g.). It is adapted to take values of $\chi$ as input or produce them as output to apply the methods presented in this paper.

\subsection{Bins and slopes of the distribution}\label{AppIshinisanBins}

The initial size distribution of grains is generated by the code according to the user's preference and is divided into bins of equal widths in the log-space. If $a_\mathrm{min}$ and $a_\mathrm{max}$ are the minimum and maximum size of a distribution with $n$ bins, the maximum and minimum size of the bin $i$, $a_{i+}$ and $a_{i,-}$, are 
\begin{align}
  a_{i,-}&=a_\mathrm{min}\zeta^{i-1},\\
  a_{i,+}&=a_\mathrm{min}\zeta^{i},
\end{align}
where $\zeta=(a_\mathrm{max}/a_\mathrm{min})^{1/n}$ is the logarithmic increment between successive bins. Similarly, for the mass of the bins,
\begin{align}
  m_{i,-}&=m_\mathrm{min}\eta^{i-1},\\
  m_{i,+}&=m_\mathrm{min}\eta^{i},
\end{align}
with $\eta=(m_\mathrm{max}/m_\mathrm{min})^{1/n} = \zeta^3$, and $m=4/3\pi \rho_\mathrm{s} a^3$ assuming spherical grains.
For a power-law distribution of the form

\begin{equation}
    \frac{\mathrm{d}n}{\mathrm{d}a} \propto a^\lambda
\end{equation}
($\lambda=-3.5$ for the MRN distribution), the initial density in each bin $i$ is

\begin{equation}
 \rho_i= \rho_g d \frac{a_{i,+}^{\lambda+4}-a_{i,-}^{\lambda+4}}{a_\mathrm{max}^{\lambda+4}-a_\mathrm{min}^{\lambda+4}},
\end{equation}
where $\rho_g$ is the gas density and $d$ is the dust-to-gas mass ratio.

If we define $\beta$ as the slope in mass of the distribution

\begin{equation}
    \frac{\mathrm{d}n}{\mathrm{d}m} \propto m^\beta,
\end{equation}
then $\beta=(\lambda-2)/3$. The slope locally changes with time, and we assume that it is a power law with a constant coefficient across the width of the bin. It is calculated as the harmonic mean of the slope on each side of the bin,

\begin{align}
    \delta_+ &= \log\left(\frac{1}{\eta}\frac{\rho_{i+1}}{\rho_i}\right), \\
    \delta_- &= \log\left(\frac{1}{\eta}\frac{\rho_{i}}{\rho_{i-1}}\right), \\
    \beta_i &= \frac{1}{\log(\eta)}\frac{2 \delta_- \delta_+}{\delta_- + \delta_+} -1, \\
    \lambda_i &= 3\beta_i + 2,
\end{align}
with the special cases

\begin{align}
    \beta_1 &= \frac{\delta_+}{\log(\eta)}, \\
    \beta_n &= \frac{\delta_-}{\log(\eta)}, \\
    \beta_i &= \frac{\delta_+}{\log(\eta)}~~~ \mathrm{if}~\delta_- \delta_+ < 0.
\end{align}

The equivalent size, cross section, and mass of each are calculated at each time step using the first, second, and third moments of the distribution in each bin. The $k$th moment, for $k=1,2,3$, reads

\begin{align}
    \langle a^k \rangle_i & = \frac{\int_{a_{i,-}}^{a_{i,+}} a^k \mathrm{d}n}{\int_{a_{i,-}}^{a_{i,+}} \mathrm{d}n}  = a_{i,-}^k \frac{\lambda_i +1}{\lambda_i +1 + k}\frac{\zeta^{\lambda_i+ 1 + k}-1}{\zeta^{\lambda_i+1}-1},\\
\end{align}
with the special case

\begin{equation}
    \langle a^k \rangle_i = a_{i,-}^k \frac{\log(\zeta)}{1-\zeta^{-k}},
\end{equation}
if $\lambda_i + 1 + k = 0$ (or is close to zero, to avoid numerical errors).

We then have $\langle a^1 \rangle_i$ the average grain size of the bin and $\langle m \rangle_i = 4/3 \pi \rho_\mathrm{s} \langle a^3 \rangle_i$ its average mass.
\begin{equation}
 n_i = \int_{a_{i,-}}^{a_{i,+}} \mathrm{d}n = \frac{\rho_i}{\langle m \rangle_i},
\end{equation}
is the number density of grains in bin $i$.

\subsection{Coagulation between two bins}

We consider two bins $i$ and $j$ ($i \geq j$) that coagulate. In the intermediate coupling regime of \citet{2007A&A...466..413O}, their coagulation rate is

\begin{equation}
    \left(\frac{\mathrm{d}n}{\mathrm{d}t}\right)_{i,j} = n_i n_j K(m_i,m_j) = n_i n_j \sqrt{\frac{8}{3\pi}}\pi (a_i+a_j)^2 \Delta V,
\end{equation}
where
\begin{equation}
    \Delta V = \left(\frac{3}{\sqrt{8}} z_0 [k_\mathrm{B}G]^{1/2} \frac{\gamma \rho_\mathrm{s}}{\mu m_\mathrm{H}} \right)^{\frac{1}{2}} a^{\frac{1}{2}} (\nh T)^{-\frac{1}{4}}.
\end{equation}
If $i=j$, the kernel has to be divided by 2 so as to not count the same grains twice.

The product of coagulation is distributed over two bins $C$ and $C+1$, with

\begin{align}
    C &= \left\lfloor  \frac{1}{\log\eta}\log\left(\frac{m_{i,-}+m_{j,-}}{m_\mathrm{min}}\right) +1\right\rfloor.
\end{align}
If we assume $m_C \ll m_j$, then the coagulation between any grain of bin $j$ and a grain of bin $i$ with a mass lower than $m_{C,-}-m_j$ produces a grain in bin $C$. The fraction $\upsilon$ of grains transferred to bin $C$ is therefore given by

\begin{equation}\label{EqFraction}
    \upsilon = \frac{\int_{m_{j,-}}^{m_{j,+}} \int_{m_{i,-}}^{m_{C,+}-m_j} (m+m') K(m,m') m^{\beta_i} m'^{\beta_j} \mathrm{d}m \mathrm{d}m'}{\int_{m_{j,-}}^{m_{j,+}} \int_{m_{i,-}}^{m_{i,+}} (m+m') K(m,m') m^{\beta_i} m'^{\beta_j} \mathrm{d}m \mathrm{d}m'}.
\end{equation}
However, if $m_{C,-}-m_j < m_{i,-}$ , then the product of coagulation is too massive for bin $C$ and $\upsilon=0$. Conversely, if $m_{C,-}-m_j > m_{i,+}$ then $\upsilon=1$.
The integrals of equation \ref{EqFraction} need to be calculated analytically due to the computational cost of performing a numerical integration for each pair of bins at each time step. In particular, if the distribution is steep, one end of the bin may not be captured accurately, yielding large errors. For coagulation kernels that are in the form  $K(m,m')=m^\alpha m'^{\alpha'}$, the analytical calculation is tedious but straightforward. Fortunately, this is the case for the kernel used in this paper (see equations \ref{EqKernel} and \ref{Eqdeltv}),

\begin{equation}\label{EqAppkernel}
  K(m,m') \propto \left(m^{\frac{1}{3}} + m'^{\frac{1}{3}}\right)^2 m^\frac{1}{6}.
\end{equation}

The analytical expression of $\upsilon$ is a function of $m^\beta$, which can lead to inaccuracies due to numerical truncation errors where the distribution is steep ($|\beta|$ is large). Fortunately, for high values of $|\beta|$, some terms can be simplified. This is not the case for specific lower values (if we have to integrate $m^{-1}$ into a log function instead of a power-law), however, so that the simplification cannot be generalized to all $\beta$. We also have to ensure that the simplification can be applied to both the top and bottom of the fraction. We detail here the procedure used in Ishinisan to minimize numerical inaccuracy.

We define the following functions:
\begin{equation}\label{EqQmmsu}
    q(m_-,m_+,\beta,u) = \left\{ \begin{array}{lr} 
    \log\left(\frac{m_+}{m_-}\right)                  & \mathrm{if}~|\beta+u|<10^{-10},\\
     & \\
    \frac{\left(\frac{m_+}{m_-}\right)^{\beta+u} -1}{\beta+u}       & \mathrm{if}~\beta<-5 ,\\
    & \\
    m_-^{\beta+u}\frac{\left(\frac{m_+}{m_-}\right)^{\beta+u} -1}{\beta+u}  & \mathrm{else},
    \end{array}\right.
\end{equation}
and  $r(m_{1,-},m_{1,+},m_{2,-},m_{2,+},\beta_1,\beta_2)$, so that
\begin{equation}\label{EqfracR}
    \upsilon = \frac{r(m_{i,-},m_{C,+}-m_j,m_{j,-},m_{j,+},\beta_i,\beta_j)}{r(m_{i,-},m_{i,+},m_{j,-},m_{j,+},\beta_i,\beta_j)}.
\end{equation}
We therefore have for the turbulent kernel 

\begin{equation}
 \begin{aligned}
 &r(m_{1,-},m_{1,+},m_{2,-},m_{2,+},\beta_1,\beta_2) =\\
 &\kappa_1 ^{\frac{5}{3}} \kappa_2^{\frac{0}{3}} q\left(m_{1,-},m_{1,+},\beta_1,\frac{7}{6}+\frac{5}{3}\right) q\left(m_{2,-},m_{2,+},\beta_2,1+\frac{0}{3}\right)\\
 +&\kappa_1 ^{\frac{4}{3}} \kappa_2^{\frac{1}{3}} q\left(m_{1,-},m_{1,+},\beta_1,\frac{7}{6}+\frac{4}{3}\right) q\left(m_{2,-},m_{2,+},\beta_2,1+\frac{1}{3}\right)\\
 +&\kappa_1 ^{\frac{3}{3}} \kappa_2^{\frac{2}{3}} q\left(m_{1,-},m_{1,+},\beta_1,\frac{7}{6}+\frac{3}{3}\right) q\left(m_{2,-},m_{2,+},\beta_2,1+\frac{2}{3}\right)\\
 +&\kappa_1 ^{\frac{2}{3}} \kappa_2^{\frac{3}{3}} q\left(m_{1,-},m_{1,+},\beta_1,\frac{7}{6}+\frac{2}{3}\right) q\left(m_{2,-},m_{2,+},\beta_2,1+\frac{3}{3}\right)\\
 +&\kappa_1 ^{\frac{1}{3}} \kappa_2^{\frac{4}{3}} q\left(m_{1,-},m_{1,+},\beta_1,\frac{7}{6}+\frac{1}{3}\right) q\left(m_{2,-},m_{2,+},\beta_2,1+\frac{4}{3}\right)\\
 +&\kappa_1 ^{\frac{0}{3}} \kappa_2^{\frac{5}{3}} q\left(m_{1,-},m_{1,+},\beta_1,\frac{7}{6}+\frac{0}{3}\right) q\left(m_{2,-},m_{2,+},\beta_2,1+\frac{5}{3}\right),
 \end{aligned}
 \end{equation}
 where $\kappa_k=m_{k,-}$ if $s_k<-5$, and $\kappa_k=1$ else, for $k=1,2$, to match the criteria of function $q$ (\ref{EqQmmsu}).

When the value of $\upsilon$ is known, the variations rates of mass density in the four involved bins are

\begin{align}
    \frac{\Delta \rho_i}{\Delta t} &= -m_i \left(\frac{\mathrm{d}n}{\mathrm{d}t}\right)_{i,j},\\
    \frac{\Delta \rho_j}{\Delta t} &= -m_j \left(\frac{\mathrm{d}n}{\mathrm{d}t}\right)_{i,j},\\
    \frac{\Delta \rho_C}{\Delta t} &= \upsilon (m_i+m_j) \left(\frac{\mathrm{d}n}{\mathrm{d}t}\right)_{i,j},\\
    \frac{\Delta \rho_{C+1}}{\Delta t} &= (1-\upsilon) (m_i+m_j) \left(\frac{\mathrm{d}n}{\mathrm{d}t}\right)_{i,j}.
\end{align}

After summing all the variation rates, we chose the time-step $\Delta t$ so that the relative variation of mass in each bin does not exceed a number $\epsilon_\mathrm{prec}$, typically smaller than one, to ensure numerical accuracy, 

\begin{equation}
    \frac{1}{\rho_i}\left(\frac{\Delta \rho_i}{\Delta t}\right)_\mathrm{tot}\Delta t < \epsilon_\mathrm{prec}.
\end{equation}
The densities are then updated
\begin{equation}
    (\rho_i)_\mathrm{new} = \rho_i + \left(\frac{\Delta \rho_i}{\Delta t}\right)_\mathrm{tot}\Delta t,
\end{equation}
and the new slopes, average size, and average mass of bins are calculated according to the calculations detailed in section \ref{AppIshinisanBins}.

\subsection{Tests}

We tested the coagulation algorithm using two simple kernels, namely the constant and additive kernels,
\begin{align}
    K_\mathrm{c}(m,m') &= 2, \\
    K_\mathrm{a}(m,m') &= m+m'.
\end{align}
\citet{Menonpego} showed that these two kernels possess self-similar solutions,

\begin{align}
    \frac{\mathrm{d}\log \rho_\mathrm{c}}{\mathrm{d}\log m} & = \left(\frac{m}{t}\right)^2 \mathrm{e}^{-\frac{m}{t}},\\
    \frac{\mathrm{d}\log \rho_\mathrm{a}}{\mathrm{d}\log m} & = \sqrt{\frac{m\mathrm{e}^{-2t}}{2\pi}} \mathrm{e}^{-m\mathrm{e}^{2t}/2}.
\end{align}
The mass density in each bin $i$ is given by 

\begin{align}
  \rho_{\mathrm{c},i} &= \left\{\begin{array}{lr}
  \left(1+\frac{m_{i,-}}{t}\right)\mathrm{e}^{-\frac{m_{i,-}}{t}} - \left(1+\frac{m_{i,+}}{t}\right)\mathrm{e}^{-\frac{m_{i,+}}{t}} & \mathrm{for}~\frac{m_{i,-}}{t}<1,\\
  \mathrm{e}^{-\frac{m_{i,-}}{t}}\left[\left(1+\frac{m_{i,-}}{t}\right) - \left(1+\frac{m_{i,+}}{t}\right)\mathrm{e}^{\frac{-m_{i,+}+m_{i,-}}{t}}\right] & \mathrm{else},
  \end{array}
  \right. \label{EqConkernel}\\
  \rho_{\mathrm{a},i} &= \left\{\begin{array}{lr}
  \mathrm{erf}\left(\frac{m_{i,+}\mathrm{e}^{-2t}}{2}\right) -  \mathrm{erf}\left(\frac{m_{i,-}\mathrm{e}^{-2t}}{2}\right) & \mathrm{for}~m_{i,-}\mathrm{e}^{-2t}<1,\\
  -\mathrm{erfc}\left(\frac{m_{i,+}\mathrm{e}^{-2t}}{2}\right) +  \mathrm{erfc}\left(\frac{m_{i,-}\mathrm{e}^{-2t}}{2}\right) & \mathrm{else},
  \end{array}
  \right.\label{EqAddkernel}
\end{align}
where $\mathrm{erf}$ and $\mathrm{erfc}$ are the error function and the complementary error function. In each case, the two formulae are mathematically equivalent but are written in this form to avoid numerical truncation errors.

The fraction $\upsilon$ has a similar form as the turbulent kernel presented in the previous section (equation \ref{EqfracR}). The function $q$ is the same, and function r  for the constant kernel reads
\begin{equation}
 \begin{aligned}
 &r_\mathrm{c}(m_{1,-},m_{1,+},m_{2,-},m_{2,+},\beta_1,\beta_2) =\\
 &\kappa_2 q\left(m_{1,-},m_{1,+},\beta_1,1\right) q\left(m_{2,-},m_{2,+},\beta_2,2\right)\\
 +&\kappa_1 q\left(m_{1,-},m_{1,+},\beta_1,2\right) q\left(m_{2,-},m_{2,+},\beta_2,1\right),
 \end{aligned}
 \end{equation}
 and for the additive kernel
\begin{equation}
 \begin{aligned}
 &r_\mathrm{a}(m_{1,-},m_{1,+},m_{2,-},m_{2,+},\beta_1,\beta_2) =\\
 &\kappa_1^2 q\left(m_{1,-},m_{1,+},\beta_1,3\right) q\left(m_{2,-},m_{2,+},\beta_2,1\right)\\
 +&2\kappa_1 \kappa_2 q\left(m_{1,-},m_{1,+},\beta_1,2\right) q\left(m_{2,-},m_{2,+},\beta_2,2\right)\\
 +&\kappa_2^2 q\left(m_{1,-},m_{1,+},\beta_1,1\right) q\left(m_{2,-},m_{2,+},\beta_2,3\right).
 \end{aligned}
 \end{equation}

\begin{figure}
\begin{center}
\includegraphics[trim=1.5cm 0.5cm 1cm 1cm, width=0.49\textwidth]{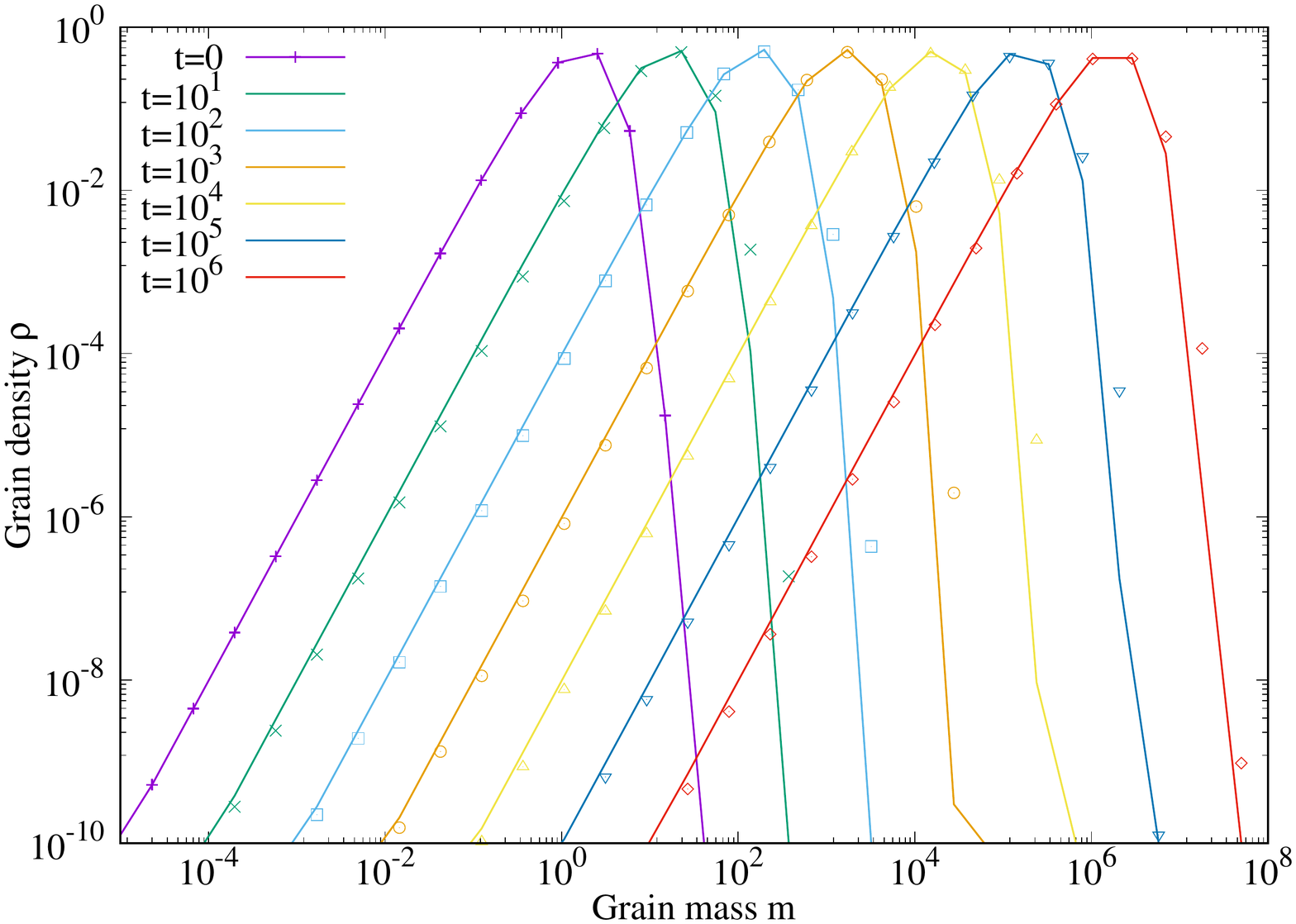}
\includegraphics[trim=1.5cm 0.5cm 1cm 1cm, width=0.49\textwidth]{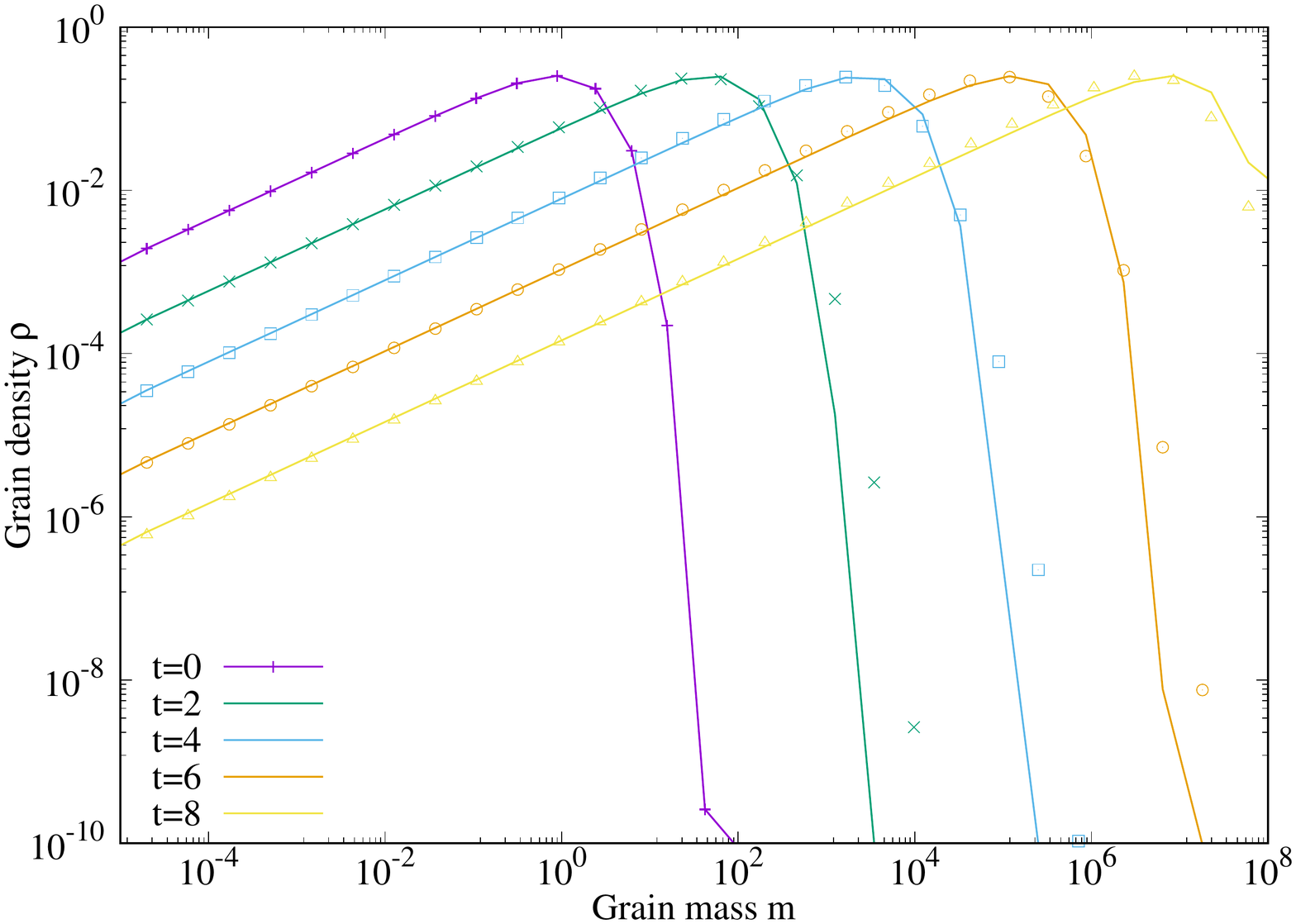}
  \caption{Test runs of Ishinisan for the constant kernel (top) and addivite kernel (bottom). The points are the simulation results and the solid lines represent the analytical solution.}
  \label{FigTestishinisan}
\end{center}
\end{figure}

We used Ishinisan to reproduce these self-similar solutions. We set up 30 bins between $m_\mathrm{min}=4\times 10^{-6}$ and $m_\mathrm{max}=4\times 10^{8}$. The initial size distribution was given by the analytical solutions (\ref{EqConkernel}) and (\ref{EqAddkernel}) at $t=0$. Figure \ref{FigTestishinisan} shows the numerical results for both kernels alongside their analytical solutions (\ref{EqConkernel}) and (\ref{EqAddkernel}). Ishinisan shows an excellent agreement with the theoretical prediction. The largest errors appear at high masses where the distribution is very steep, which affects only a minor fraction of the total mass of the grains.
 
\bibliographystyle{aa}
\bibliography{MaBiblio}

\end{document}